\documentclass[aps,pra,reprint,superscriptaddress,floatfix]{revtex4-1}

\usepackage{bm}
\usepackage{amsmath,amssymb,cases}
\usepackage{graphicx,color}
\usepackage[colorlinks=true,urlcolor=blue]{hyperref}

\begin{document}

\title{Dynamical generation of dark-bright solitons through the domain wall of two immiscible Bose-Einstein condensates}

\author{Maria Arazo}
\affiliation{Departament de F\'isica Qu\`antica i Astrof\'isica,
Facultat de F\'{\i}sica, Universitat de Barcelona, 08028
Barcelona, Spain}
\affiliation{Institut de Ci\`encies del Cosmos de la Universitat de
Barcelona, ICCUB, 08028 Barcelona, Spain}
\author{Montserrat Guilleumas}
\affiliation{Departament de F\'isica Qu\`antica i Astrof\'isica,
Facultat de F\'{\i}sica, Universitat de Barcelona, 08028
Barcelona, Spain}
\affiliation{Institut de Ci\`encies del Cosmos de la Universitat de
Barcelona, ICCUB, 08028 Barcelona, Spain}
\author{Ricardo Mayol}
\affiliation{Departament de F\'isica Qu\`antica i Astrof\'isica, Facultat
de F\'{\i}sica, Universitat de Barcelona, 08028 Barcelona, Spain}
\affiliation{Institut de Ci\`encies del Cosmos de la Universitat de
Barcelona, ICCUB, 08028 Barcelona, Spain}
\author{Michele Modugno}
\affiliation{Department of Physics, University of the Basque Country UPV/EHU, 48080 Bilbao, Spain}
\affiliation{IKERBASQUE, Basque Foundation for Science, 48013 Bilbao, Spain}

\date{\today}

\begin{abstract}
We theoretically investigate the one-dimensional dynamics of a dark soliton in a two-component immiscible mixture of Bose-Einstein condensates with repulsive interactions. We analyze the reflection and transmission of a soliton when it propagates through the domain wall, and we show that a  dark-bright soliton can be dynamically generated by the interaction of the dark soliton with the domain wall, outside the regime of parameters where stationary solutions are known to exist. The dynamics of this dark-bright soliton is harmonic like, with a numerical frequency that is in good agreement with the predictions of a semi-analytical model.
\end{abstract}

\maketitle

\section{Introduction}

Solitons are localized, nondispersive excitations that can transport energy and momentum in a nonlinear medium \cite{pitaevskii2016}. They are topological states that propagate keeping their shape unaltered, as a result of the competition between dispersion and the nonlinearity of the system.

The experimental flexibility and high level of control of Bose-Einstein condensates (BECs), such as dimensionality and strength of the interatomic interactions, have led this system to be an excellent playground for the study of matter-wave solitons and topological excitations \cite{frantzeskakis2010,kevrekidis2010}. Solitonic states in BECs can be investigated within a mean-field description, by means of the Gross-Pitaevskii (GP) equation. It incorporates a nonlinear term that takes into account the interatomic interaction. Depending on the sign of the latter, two types of matter waves can be found in a single BEC: dark or bright solitons, for  repulsive and attractive interactions, respectively.

Multicomponent condensates with different intra- and
interspecies interactions offer the possibility to investigate new families of solitonic states in different regimes, in which the two components are miscible or immiscible~\cite{Pu1998,Tim1998}. Such two-component BECs can be experimentally produced from two different hyperfine states of the same atomic species, for instance $^{87}$Rb~\cite{Myatt1997,Bec2008}, or from two different atomic species~\cite{Hall1998}. Depending on the ratio between the interaction constants, new solitonic configurations have been experimentally realized, for instance, dark-dark solitons~\cite{Hoe2011} or dark-bright solitons~\cite{Mid2011,Bec2008,Ham2011}. The latter structure is specially appealing because the bright component, with repulsive intraspecies interaction, can exist because the density depletion of the dark component plays the role of an effective confining potential. The dark soliton in one component hosts the atoms of the bright one, as in vortex-bright soliton configurations. The latter are topological states formed by vortices with massive cores~\cite{And2000,Law2010,Ric2020}, one component supports a quantized vortex, and the other fills the core.

Dark-bright (DB) solitons are exact solutions in the two-component one-dimensional (1D) Manakov limit, where all the interaction constants are equal (see Ref.~\cite{Kev2016} and references therein). For the general case (\textit{i.e.}, non Manakov), explicit analytic solutions for DB solitons have been obtained in Ref.~\cite{Yan2015} for a restricted range of the interactions; there it has also been shown that other solutions (\textit{e.g.}, DB soliton trains) can still be found numerically even beyond those limits. Many other features of DB solitary waves have been investigated in the literature. For example, the dynamics of a DB soliton in a harmonic trap, whose oscillation frequency is smaller than the one of a dark soliton in a single-component (owing to the presence of the massive core of the bright filling component that slows down the oscillation) ~\cite{busch2001,Bec2008,Mid2011}; DB soliton trains generated by the counterflow of two components~\cite{Ham2011}; collision between a dark and a DB soliton~\cite{Bec2008}, as well as scattering of a DB soliton by an impurity~\cite{Ach2011,Alo2019}.

In this article we shall consider an immiscible two-component system, whose equilibrium state is characterized by the phase separation of the two components, each in a different domain. The interface region between the two components is the so-called \emph{domain wall}. The propagation of an imprinted dark soliton in two immiscible BECs has been previously investigated in 2001 by \"Ohberg and Santos~\cite{Ohb2001,Ohb2001b}, and more recently in Ref.~\cite{Zhe2019}, for a fixed set of the parameters.  The aim of this paper is to investigate comprehensively the reflection and transmission of nonlinear matter waves in two immiscible BECs. In particular we want to study the effect of the domain wall in a wider range of interaction parameters in the immiscibility regime. We consider general interaction coefficients, motivated by the tunability of the scattering lengths by means of Feshbach resonances. Since the shape and features of a domain wall depend on the interparticle interactions, one can expect different dynamical behaviors when the moving soliton encounters the domain wall. We show that a DB soliton can be dynamically generated after the reflection and transmission of a dark soliton through a domain wall of two immiscible condensates. We have found that these DB solitons are dynamically generated in a region of interaction parameters where static solutions of this type cannot be obtained~\cite{Yan2015}.

This paper is organized as follows. In Sec.~\ref{sec:system}, we introduce the system and the theoretical framework, based on the mean-field GP theory. In Sec.~\ref{sec:numerics} we present the numerical results obtained by solving the two-coupled time-dependent GP equations by varying the interspecies interaction within the immiscibility regime, for fixed intraspecies interactions. We analyze the reflection or transmission of the initially imprinted dark soliton. We show that the interaction of the moving dark soliton with the domain wall generates a DB soliton for a wide range of intraspecies interactions in the immiscibility regime. In Sec.~\ref{sec:dynamics}, we study the DB soliton dynamics and provide a semi-analytical expression for its harmonic frequency that is in good agreement with the numerical one. To sum up, we present our conclusions and perspectives for future work in Sec.~\ref{sec:conclusions}.

\section{The system}\label{sec:system}

We consider a two-component Bose-Einstein condensate, confined in a highly elongated harmonic potential. The longitudinal $(\omega_x)$ and transversal $(\omega_\perp)$ frequencies are such that $\omega_x \ll  \omega_\perp$. In the mean-field regime, the system can be  accurately described by the following 1D two-coupled GP equations in dimensionless units:
\begin{equation}
i\hbar \frac{\partial {\psi}_{i}}{\partial {t}}=\left[-\frac12\frac{\partial^2}{\partial {x}^2} + \frac12{x}^{2}
+ {g}_{ii} |{\psi}_{i}|^{2}
+ {g}_{12}|{\psi}_{j}|^{2}\right]{\psi}_{i} \,,
\label{1D-GP-eq}
\end{equation}
where $\psi_i(x,t)$ $(i=1,2)$ denote the mean-field wave functions of the two components, normalized to one. We have used the longitudinal trap length, $a_x=\sqrt{\hbar/(m \omega_x)}$, as unit length, $\hbar \omega_x$ as unit energy, and $t_x=1/\omega_x$ as unit time. The effective 1D dimensionless coupling constants are:
$$g_{ii}= 2 N \, \frac{\omega_\perp}{\omega_x} \frac{a_i}{a_x} \,,
\quad
{g}_{12}= 2 N \, \frac{\omega_\perp}{\omega_x} \frac{a_{12}}{a_x} \,, $$
with $a_i$ and $a_{12}$, the intraspecies and interspecies scattering lengths, respectively, and $N$ the number of atoms. Here, for the sake of conceptual clarity, we assume that the intraspecies interaction is the same for both components, $a \equiv a_{1}=a_{2}$, and therefore $g \equiv g_{11}=g_{22}$. Moreover, we consider repulsive interactions, such that the immiscibility condition~\cite{Tim1998,Pu1998}, $a_{12} > \sqrt{a_{1}a_{2}}\,$, is fulfilled: $g_{12} > g > 0$.

Owing to the above conditions, the system is prepared with the component $1$ on the left (L) side of the trap, and the other component ($2$) at its right (R). In the following, we may indicate the two components equivalently as $i=1,2$ or $L/R$. Initially, the domain wall lies at the trap center $(x=0)$. We will see below that its exact shape and position is affected by the value of $g_{12}/g$, as well as by the presence and dynamics of a moving soliton. One should bear in mind that despite the immiscibility condition, a minority fraction of the $1$-component coexists also on the right side of the trap, and vice versa for the $2$-component. This small overlap of the two components in the tiny region around the domain wall is indeed crucial for the dynamical generation of DB solitons, as we will discuss later on.

As initial state, we consider a dark soliton imprinted at rest in the right component $(i=2)$, and located at $x_0$. The system can be described by the following ansatz \cite{tsuzuki1971,pitaevskii2016}:
\begin{eqnarray}
\psi_1(x)&=&\psi_1^{\rm{gs}}(x) \,, \nonumber \\
\psi_2(x)&=&\psi_2^{\rm{gs}}(x) \tanh\left(\frac{x-x_{0}}{\sqrt{2} \, \xi}\right) \,,
\label{eq:ansatz}
\end{eqnarray}
where $\psi_{1}^{\rm gs}(x)$ and $\psi_2^{\rm gs}(x)$ are the ground state solutions of the two components of the immiscible mixture in the elongated trap,
$\xi=1/\sqrt{2 \mu_0}$
is the (dimensionless) healing length, and $\mu_0$ is the (dimensionless) chemical potential of the uniform singly-component condensate with density $n_0 \equiv |\psi_2^{\rm gs}(x_0)|^2$.
Numerically, the initial state is prepared by letting evolve the trial wave function (\ref{eq:ansatz}) in imaginary time
\footnote{The initial soliton is imprinted at $x=x_{0}$ as follows. For each value of $g_{12}$ we generate the ground state of the system by a standard steepest descent algorithm (that is, an imaginary time evolution). Then, we redefine the right component by multiplying it by a soliton profile, as in Eq. (\ref{eq:ansatz}). Finally, we let again evolve in imaginary time the new trial wave function, until we reach the desired tolerance in the solution of the stationary GP equation \cite{modugno2003}.}.

\section{Numerical results}
\label{sec:numerics}

In order to perform the numerical calculations, we fix the dimensionless intraspecies interaction to $g= 3 \times 10^3$. It gives the order of magnitude for a typical condensate in the mean-field regime. For instance, a $^{87}$Rb BEC with $N=10^5$ atoms, confined in a tight transverse harmonic trap with frequencies $\omega_x= 2 \pi \times  10\,$Hz, $\omega_\perp =   2 \pi \times100\,$Hz, and intraspecies $s$-wave scattering length $a \simeq 100 \, a_0$, with $a_0$ being the Bohr radius.
To illustrate the different possible scenarios, we investigate the soliton dynamics for different values of the interspecies interaction with $g_{12}/g\in(1,4],$ in the immiscibility regime.

\begin{figure}[ht]
\includegraphics[width=0.99\columnwidth]{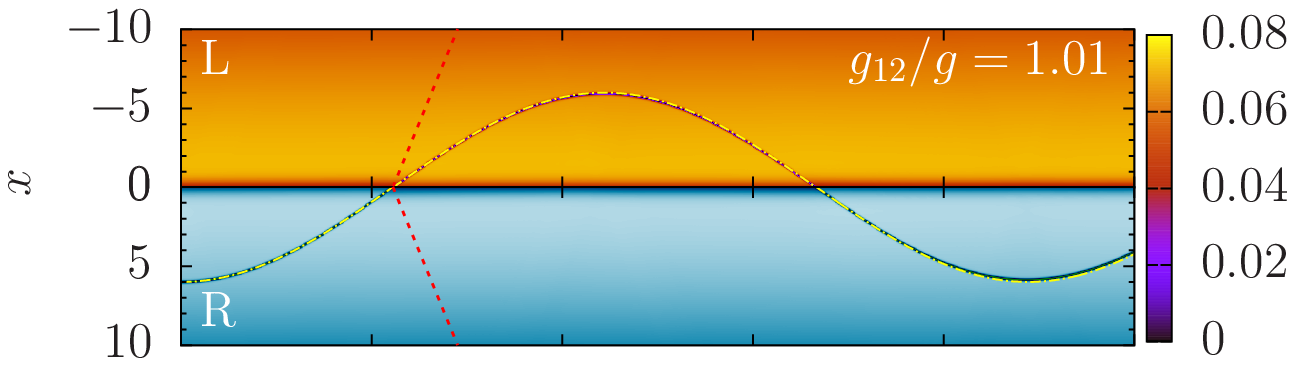}
\includegraphics[width=0.99\columnwidth]{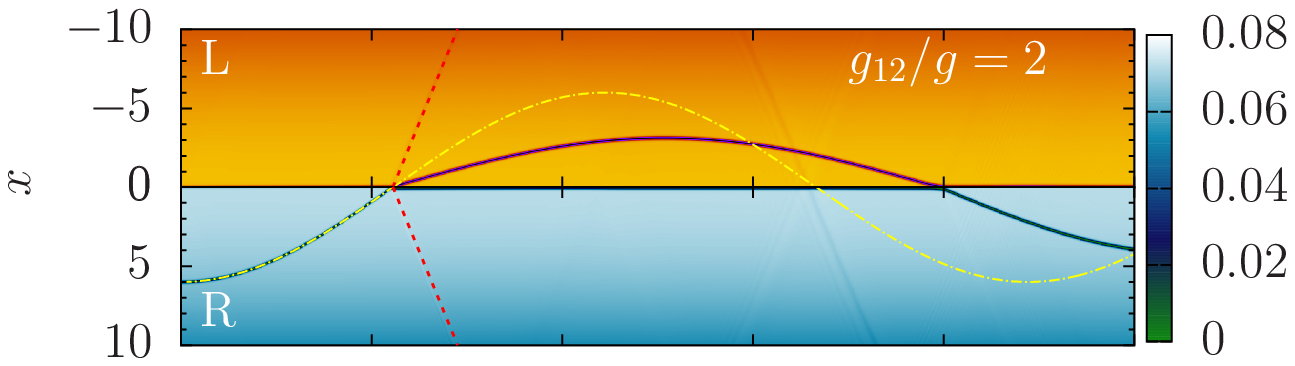}
\includegraphics[width=0.99\columnwidth]{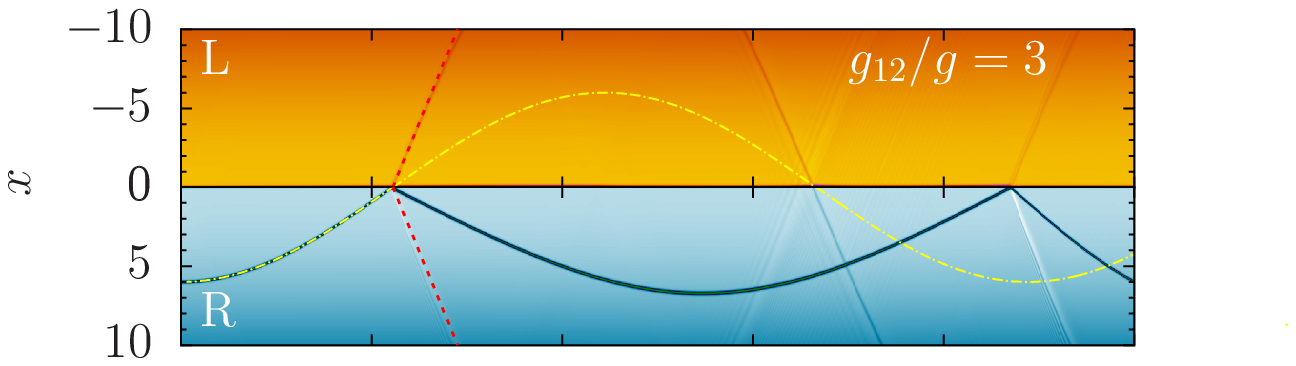}
\includegraphics[width=0.99\columnwidth]{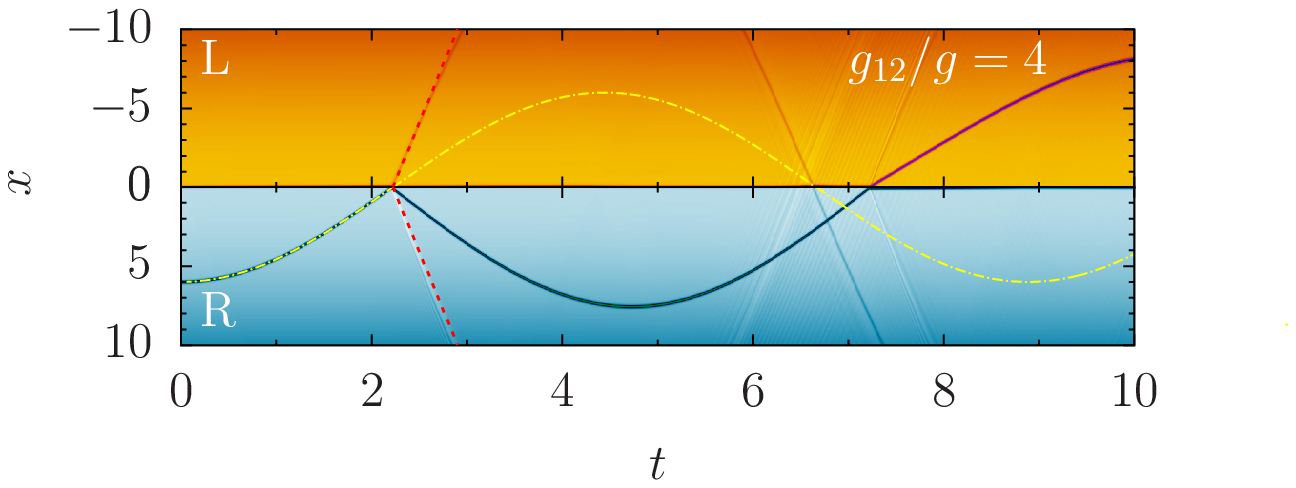}
\caption{
Evolution of the soliton density obtained from the numerical solution of the GPE (\ref{1D-GP-eq}). From top to bottom: $g_{12}/g=1.01 ,2,3$ and $4$. The point-dashed yellow line represents the oscillation of a dark soliton in a single component, with frequency $\omega_{0}=1/\sqrt{2}$ \cite{pitaevskii2016,konotop2004}. Straight lines represent phonon trajectories traveling at the speed of sound (only those generated at the first encounter with the domain wall are highlighted in red, see text).
}
\label{fig1:dens_x_t}
\end{figure}
Initially, we imprint the soliton at $x_{0}=6$, without loss of generality. Then, we let it evolve freely, according to Eq.~(\ref{1D-GP-eq}). The dark soliton acquires an initial velocity given by the local density gradient around $x_0$, which is a consequence of the harmonic confinement. The soliton moves towards the domain wall following the same trajectory as in the absence of the other component \cite{kivshar1994,konotop2004}, whose presence becomes important only from the domain wall on. This is clearly shown in Fig.~\ref{fig1:dens_x_t} where we plot, for different values of $g_{12}/g$, the evolution of the density obtained from the numerical solution of Eq.~(\ref{1D-GP-eq}).
The trajectory of the soliton, defined by the location of the density depletion as a function of time, is represented by the black curve which is sinusoidal-like in pieces (see below).

Figure~\ref{fig1:dens_x_t} shows that before reaching the domain wall, represented by the horizontal line at $x=0$, the soliton trajectory obtained from the numerical solution overlaps the harmonic dashed yellow line. The latter one, with frequency $\omega_{0}=1/\sqrt{2}$,  corresponds to an unperturbed dark soliton moving in a single harmonically confined one-dimensional (1D) BEC \cite{pitaevskii2016,konotop2004}.
When the dark soliton encounters the domain wall, we obtain two distinctive behavior depending on the interparticle strength ratio $g_{12}/g$, as shown in the top and bottom panels of Fig.~\ref{fig1:dens_x_t}.
These scenarios comprise  transmission and reflection, respectively. They are in agreement with the ones described in Refs.~\cite{Ohb2001,Ohb2001b,Zhe2019}, obtained for a fixed set of values of the interactions and different values of the initial position of the imprinted dark soliton.
\begin{figure}[t]
\includegraphics[width=0.98\columnwidth]{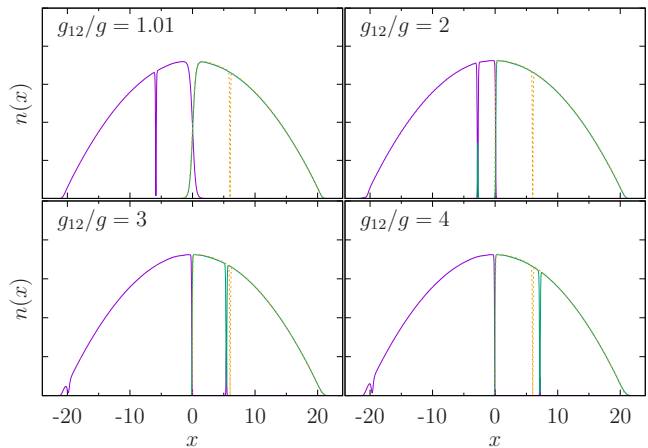}
\caption{Density snapshots at $t=4.2$ (corresponding to the squares in Fig.~\ref{fig3:soliton_trajectory}) for different values $g_{12}/g$. Solid green (magenta) line corresponds to the density of the right (left) component. The dashed yellow line represents the initial configuration of the right component, with the dark  soliton imprinted at $x_{0}=6$. The tiny density depletion at $x\simeq-20$ in the bottom panels corresponds to the phonon.}
\label{fig2:dens_snapshots}
\end{figure}
To illustrate these two different situations, we plot in Fig.~\ref{fig2:dens_snapshots} the density profiles at $t=4.2$, which correspond to snapshots after the first collision of the dark soliton with the domain wall. The panels correspond to the same values of $g_{12}/g$ as in Fig.~\ref{fig1:dens_x_t}. The initial density profile with the imprinted dark soliton at $x_{0}=6$ in the right component is also depicted as a dashed yellow line.
The top panels of Fig.~\ref{fig2:dens_snapshots} show the transmission of the soliton from the right to the left component through the domain wall. When $g_{12}/g \simeq 1$ the dark soliton is just transferred to the other component.
Interestingly, when the interparticle strength slightly increases, $1.5 \lesssim g_{12}/g \lesssim 2$  (top right panel), the transmitted soliton drags atoms of the right component forming a DB soliton.

By exploring different interparticle strength values, we have seen that the number of dragged atoms inside the dark soliton (the bright component) increases with $g_{12}/g$. This dependence will be discussed later on (see also Fig.~\ref{fig5:freq}). Since the effective mass of the moving ``object'' increases when it is filled with atoms of the bright component, a DB soliton slows down with respect to a single dark soliton. Hence, this produces a decrease of the slope of the soliton trajectory as shown in Fig.~\ref{fig1:dens_x_t} (see top panels).

Increasing further the interparticle repulsion ($2.5 \lesssim g_{12}/g $), the domain wall becomes sharper and behaves as an impenetrable wall. After the collision, the dark soliton has not enough energy to be transferred, but it drags some atoms of the other component in the domain wall and it is reflected back in the initial component as a DB soliton. The bottom panels of Fig.~\ref{fig2:dens_snapshots} show the density snapshots of the reflected DB solitonic state. Despite the two components slightly overlap in the domain wall, due to the large interparticle repulsion, the density depletion of the reflected soliton generates an attractive effective potential that drags some atoms of the left component.

Remarkably, these DB solitons are dynamically generated outside the regime of parameters where explicit analytical solutions are known to exist, namely for $g_{12}>\max(g_{11},g_{22})$ \cite{frantzeskakis2010,Yan2015}. They are dynamically created by the interaction of the moving dark soliton with the domain wall, and they have not been previously observed for the range of parameters used in Refs.~\cite{Ohb2001,Ohb2001b,Zhe2019}.

We have also checked that the same dynamical behaviors appear regardless of the initial position of the soliton, which only affects the values of $g_{12}/g$ characterizing the different dynamical regimes. For example, for $x_0=2$ the  transition between the transmission and reflection regimes occurs at smaller values of the interparticle strength ($g_{12}/g\simeq1.14$) because of the lower soliton velocity.

In Fig.~\ref{fig1:dens_x_t} are also evident shallow density depletions (light gray straight lines) that appear and propagate after the soliton is transferred or reflected at the domain wall. They correspond to the emission of phonons traveling at the speed of sound, $c=\sqrt{g \, n_0}$ \cite{Mun2015}. We have marked the first phonon trajectories as dashed red lines. The density modulation corresponding to the phonon excitation appears clearly close to the left boundary in the density profiles of the bottom panels of Fig.~\ref{fig2:dens_snapshots}.
When the phonon excitation reaches the condensate boundary it is reflected back towards the center of the system. Oblique gray lines indicate phonon excitations that propagate in each component from left to right (negative slope) and from right to left (positive slope). It is worth noting that the soliton speed (line slope) is slower than the speed of sound (phonon speed).

After the first interaction with the domain wall, the soliton travels towards the boundary of the system backwards to the interface, slowing down its velocity. As in a harmonic motion, the DB soliton stops and then travels back towards the domain wall. This is clearly shown in the harmonic-like soliton trajectories in Fig.~\ref{fig1:dens_x_t}.

\begin{figure}[t]
\includegraphics[width=0.98\columnwidth]{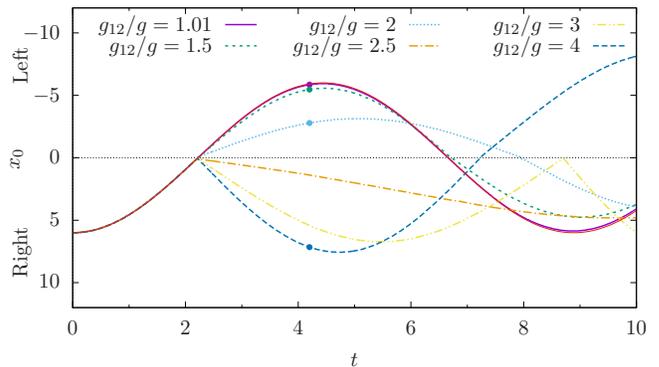}
\caption{Evolution of the center of the soliton as a function of time, for a dark soliton initially imprinted at $x_{0}=6$. The density distribution at time $t=4.2$ (corresponding to the marked squares) is shown in Fig. \ref{fig2:dens_snapshots}. The thin red line corresponds to the unperturbed trajectory of a single component dark soliton confined in a harmonic trap.}
\label{fig3:soliton_trajectory}
\end{figure}

In order to discuss the second collision with the domain wall, we show in Fig.~\ref{fig3:soliton_trajectory} the time evolution of the soliton center, $x_0(t)$, for different values of the interspecies interaction. As a reference, we also plot as a thin red line the unperturbed trajectory of a single component dark soliton confined in a 1D harmonic trap.
Close to the Manakov limit, where the intra- and interspecies interactions are equal (here $g_{12}/g=1.01$), the dark soliton transmits from one component to the other through the domain wall. After the first and second collision, the soliton trajectory in both components follows the harmonic trajectory with frequency $\omega_{0} \simeq 1/\sqrt{2}$, as discussed before.
The presence of the domain wall only produces a small perturbation of the trajectory of the transmitted dark soliton in the new component.

\begin{figure}[b]
\includegraphics[width=0.98\columnwidth]{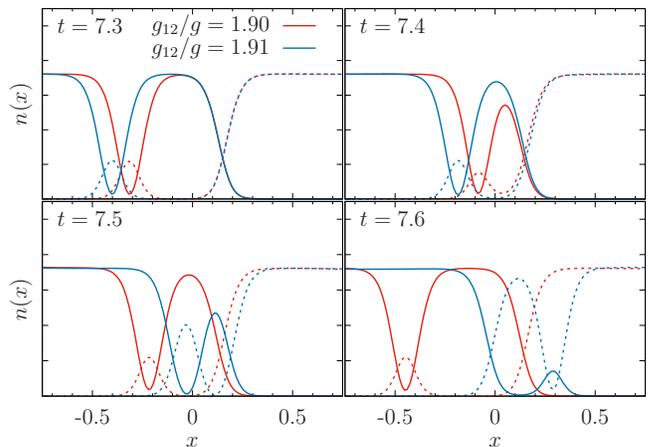}
\caption{The DB soliton travels from the left component towards the domain wall. Snapshots of the density profile close to the domain wall, for $g_{12}/g = 1.90 \, (1.91)$ depicted with red (blue) lines. Solid (dashed) lines correspond to the left (right) component.  Snapshots correspond to t=7.3, 7.4, 7.5 and 7.6. The DB soliton is reflected (transmitted) for $g_{12}/g = 1.90 \, (1.91)$. Notice that the position of domain wall, initially located at the center of the harmonic trap, slightly moves depending on the soliton position.
}
\label{fig4:domain_wall}
\end{figure}

In general, the soliton dynamics after the second collision with the domain wall follows the same behavior as after the previous collision. Namely, first the DB soliton produces emission of phonons when it interacts with the domain wall, as well as some perturbations; afterwards it is transferred or reflected. However, there are some particular cases where the perturbations generated in the domain wall substantially alter the subsequent dynamics: the soliton is transferred instead of being reflected as in the first collision (or viceversa). See, for example, the trajectory for $g_{12}/g=4$ in Fig.~\ref{fig3:soliton_trajectory}: in the first collision the soliton is reflected, whereas in the second one it is transmitted to the other component. We have verified that this behavior cannot be explained in terms of the critical velocity argument proposed in Refs.~\cite{Ohb2001,Ohb2001b}, and this suggests that the density deformations that take place in the domain wall may also play an important role, see  Fig.~\ref{fig4:domain_wall}.
In that figure we show a zoom of the density profile around the domain wall at different times, around $t\simeq7.5$, for two close values of the interaction $g_{12}/g = 1.90$ (red lines) and $g_{12}/g = 1.91$ (blue lines). The solid lines correspond to the left component, whereas the right component is represented by dashed lines. One can see that when the soliton interacts with the domain wall the latter induces a back action onto the soliton modifying its subsequent dynamics: the soliton is reflected for $g_{12}/g=1.90$, whereas it is transferred to the other component for $g_{12}/g=1.91$. We remark that the dynamics for $g_{12}/g=1.90$ is an exception in the $1.5 \lesssim g_{12}/g \lesssim 2$ range. We also mention that, in certain conditions, the domain wall can trap the soliton for some time, before it is either transmitted or reflected. This behavior is similar to that discussed in Ref.~\cite{Ohb2001b}.

\section{Dynamics of the dark-bright soliton}
\label{sec:dynamics}

\begin{figure}[h!]
\includegraphics[width=\columnwidth]{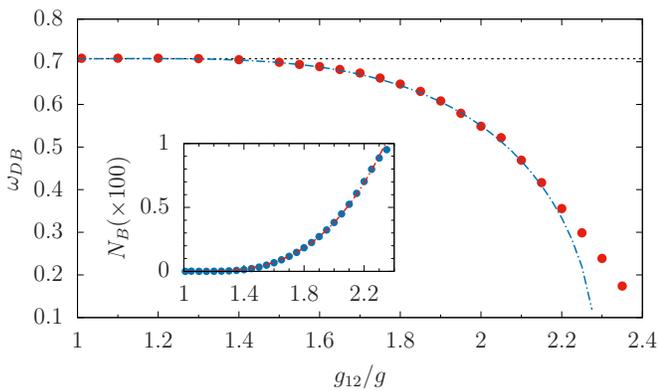}
\caption{Oscillation frequency of the DB soliton transmitted in the left component, obtained from the numerical solution of the GP equation (red points), as a function of the interspecies interaction strength $g_{12}/g$. The horizontal dashed line represents the unperturbed result $\omega_0=1/\sqrt{2}$. The dash-dotted line corresponds to the analytical prediction of Eq.~(\ref{eq:omega-DB}). Inset: rescaled number of atoms $N_B$ in the bright soliton, as a function of the interparticle strength $g_{12}/g$. The red dashed line represents a fit of the data, of the form $\alpha(g_{12}/g)^\beta$ with $\alpha\simeq3.8\times 10^{-3}$ and $\beta\simeq3.3$.}
\label{fig5:freq}
\end{figure}

As we have anticipated in the previous section, once the DB soliton has been formed at the domain wall, it starts performing a harmonic oscillation in the left component (until it gets back to the domain wall). Indeed, we have verified that the trajectory of the DB soliton core can be fitted very accurately with a sinusoidal function. Its characteristic frequency is shown in Fig.~\ref{fig5:freq}.

For $g_{12} > 1.4$ the soliton frequency starts to depart
from the unperturbed result $\omega_0=1/\sqrt{2}$ (for a dark soliton alone, horizontal line), signaling the presence of a significant drag of atoms in the bright component, which produces a slowing down of the oscillation. It is important to remark that these DB solitons are dynamically generated for $g_{12}>\max(g_{11},g_{22})$, a regime of parameters where explicit analytical solutions are not available \cite{frantzeskakis2010}. Then, in order to compare the oscillation frequency with an analytical estimate, we assume the following ansatz with the effect of the bright component being treated as a perturbation of the dark soliton frequency $\omega_0$.
In particular, we use the fact that in the Manakov case, $g_{12}=g=1$, the DB soliton frequency is given by the following expression \cite{busch2001,Yan2015}
\begin{equation}
\omega^2_{M} \simeq \frac12 \left[1 -\frac{N_{B}/g}{4\sqrt{\mu +(N_{B}/4g)^2}}\right] \,,
\end{equation}
where $N_B$ is the rescaled number of atoms in the bright soliton (we recall that the total density of each component is normalized to one)
\begin{equation}
N_B\equiv\int|\psi_{B}(x)|^{2} \,dx\,,
\end{equation}
and $\mu$ is the chemical potential. Notice also the factor $1/g$ rescaling the number of atoms in the bright soliton, which comes from the fact that in our formulation the densities are not rescaled by $g_{11}\equiv g$ as in Ref.~\cite{Yan2015}.
Then, we make an analytical continuation to $g_{12}/g>1$ assuming that the correction to $\omega_{DB}^{2}$, which depends on the interaction between the two components, has to be proportional to $g_{12}$, to lowest order.
This yields
\begin{equation}
\omega^2_{DB} \simeq \frac12 \left[1 -\frac{(g_{12}/g) N_{B}}{4\sqrt{\mu +(N_{B}/4g)^2}}\right] \,,
\label{eq:omega-DB}
\end{equation}
where $N_{B}$ also depends on $g_{12}$. The behavior of $N_{B}(g_{12})$ can be estimated by fitting the bright soliton density profile as \cite{busch2001,Yan2015}
\begin{equation}
|\psi_{B}(x)|^{2}= (\kappa \, N_{B}/2)~\text{sech}^2\left[\kappa(x-x_{0})\right],
\label{eq:bright-soliton}
\end{equation}
where $\kappa\simeq\sqrt{\mu}$ is the bright soliton width, and $x_0$  the position of the DB soliton. We notice that, once the DB soliton has been formed, both $N_{B}$ and $\kappa$ do not show any significant dependence on time, until the soliton is eventually reabsorbed at the domain wall. The behavior of $N_{B}$ as a function of $g_{12}/g$ is shown in the inset of Fig.~\ref{fig5:freq}.

This figure shows that for $g/g_{12}\gtrsim1.4$ the atoms of the left component start to fill the core of the dark soliton in the right component due to the interparticle repulsion, and this produces a slowing down of the oscillation frequency of the soliton.
Combining the above results with Eq.~(\ref{eq:omega-DB}) we obtain the semi-analytical estimate for the DB soliton frequency $\omega_{DB}$ shown in Fig.~\ref{fig5:freq} as a dashed line. Remarkably, this simple ansatz reproduces with great accuracy the frequency obtained from the numerical simulation of the GP equation, for $1\le g_{12} \lesssim2.3$.

\section{Conclusions}
\label{sec:conclusions}

In this paper we have investigated the reflection and transmission of a dark soliton through the domain wall of a 1D immiscible mixture.
We have shown that depending on the interparticle strength, a DB soliton is formed when the initially imprinted dark soliton moves across the domain wall.
Interestingly, these DB solitons are dynamically generated outside the regime of parameters where explicit analytical solutions have been demonstrated to exist.
This opens an interesting scenario for producing DB solitons in this new dynamical regime, which should be easily accessible in ultracold atom experiments~\cite{Lee2016}.
Once the DB soliton is created, it follows an harmonic-like trajectory. When it encounters the domain wall, the DB soliton can be reflected or transferred through it.
By assuming that the effect of the bright component can be treated as a perturbation, we have shown that a semi-analytical expression for the frequency of the DB soliton can be obtained by analytical continuation for $g_{12}/g\gtrsim1$ of the Manakov case discussed in Refs. \cite{And2000,Yan2015}.
Indeed, the frequency of the DB soliton oscillation obtained from the numerical solution of the GP equation is in good agreement with the predictions of the semi-analytical model.
Nonetheless, a more detailed investigation of the interaction and back-action between the domain wall and the DB soliton is required in order to shed light on the `microscopic' mechanisms that take place. This and other natural extensions of the present work, like the effect of the dimensionality of the system, are subjects that deserve further exploration and they will be presented in a future work.

\section*{Acknowledgments}

M. A., M. G. and R. M. acknowledge support from the Ministerio de Econom\'{\i}a y Competitividad (Contract No.~FIS2017-87801-P), from Ministerio de Ciencia e Innovaci\'on (Contract No.~PID2020-114626GB-I00) and from Secretaria d’Universitats i Recerca del Departament d’Empresa i Coneixement de la Generalitat de Catalunya, co-funded by the European Union Regional Development Fund within the ERDF Operational Program of Catalunya (Project QuantumCat, Ref.~001-P-001644).
M. A. acknowledges financial support from MINECO through Grant No.~PRE2018-084091.
M. M. acknowledges support from the Spanish Ministry of Science, Innovation and Universities and the European Regional Development Fund FEDER through Grant No.~PGC2018-101355-B-I00 (MCIU/AEI/FEDER, UE), and the Basque Government through Grant No.~IT986-16.

\bibliography{soliton}
\end{document}